\documentclass{nature}
\usepackage[english]{babel}
\usepackage{tabulary,graphicx,amsmath,amsfonts,amssymb}
\usepackage[T1]{fontenc}
\usepackage{ucs}
\usepackage{tikz}
\usepackage{braket}
\usepackage{pgfplots}
\usepackage{csquotes}
\usepackage{hhline}

\usepackage{dcolumn}
\usepackage{tabularx}
\usepackage{longtable}
\usepackage{float}
\usepackage{epsfig}
\usepackage{latexsym}
\usepackage{amssymb}
\usepackage{amsmath}
\usepackage{amsthm}
\usepackage{amsfonts}
\usepackage{ifthen}
\usepackage{revsymb}
\usepackage{yfonts}
\usepackage{graphicx}
\usepackage{algpseudocode}
\usepackage{bbm}
\usepackage{multirow}
\usepackage{gensymb}
\usepackage{epstopdf}
\usepackage{listings}
\usepackage{color}
\definecolor{mygreen}{rgb}{0,0.6,0}
\definecolor{mygray}{rgb}{0.5,0.5,0.5}
\definecolor{mymauve}{rgb}{0.58,0,0.82}

\usepackage{afterpage}
\usepackage{verbatim}
\usepackage{soul}
\usepackage{lineno, blindtext}
\usepackage{footnote}
\usepackage{afterpage}
\usepackage{algpseudocode}
\usepackage[normalem]{ulem}
\usepackage{algorithm}
\usepackage{algpseudocode}
\usepackage{mathtools}

\usepackage{longtable}
\usepackage{xcolor}
\definecolor{darkred}  {rgb}{0.5,0,0}
\definecolor{darkblue} {rgb}{0,0,0.5}
\definecolor{darkgreen}{rgb}{0,0.5,0}
\usepackage{amsmath}

\newcommand{\ham}{\mathcal{H}}
\makeatletter
\renewenvironment{figure}
               {\@float{figure}}
               {\end@float}
\renewenvironment{figure*}
               {\@dblfloat{figure}}
               {\end@dblfloat}
\renewenvironment{table}
               {\@float{table}}
               {\end@float}
\renewenvironment{table*}
               {\@dblfloat{table}}
               {\end@dblfloat}

\makeatother

\usepackage{url,multirow,morefloats,floatflt,cancel,tfrupee}
\makeatletter

\AtBeginDocument{\@ifpackageloaded{textcomp}{}{\usepackage{textcomp}}}
\makeatother
\usepackage{colortbl}
\usepackage{xcolor}
\usepackage{pifont}
\usepackage[nointegrals]{wasysym}
\makeatletter

\def\mcWidth#1{\csname TY@F#1\endcsname+\tabcolsep}

\def\cAlignHack{\rightskip\@flushglue\leftskip\@flushglue\parindent\z@\parfillskip\z@skip}
\def\rAlignHack{\rightskip\z@skip\leftskip\@flushglue \parindent\z@\parfillskip\z@skip}

\if@twocolumn\usepackage{dblfloatfix}\fi 
\AtBeginDocument{
\expandafter\ifx\csname eqalign\endcsname\relax
\def\eqalign#1{\null\vcenter{\def\\{\cr}\openup\jot\m@th
  \ialign{\strut$\displaystyle{##}$\hfil&$\displaystyle{{}##}$\hfil
      \crcr#1\crcr}}\,}
\fi
}

\let\lt=<
\let\gt=>
\def\processVert{\ifmmode|\else\textbar\fi}

\@ifundefined{subparagraph}{
\def\subparagraph{\@startsection{paragraph}{5}{2\parindent}{0ex plus 0.1ex minus 0.1ex}%
{0ex}{\normalfont\small\itshape}}%
}{}

\newcommand\role[1]{\unskip}
\newcommand\aucollab[1]{\unskip}
  
\@ifundefined{tsGraphicsScaleX}{\gdef\tsGraphicsScaleX{1}}{}
\@ifundefined{tsGraphicsScaleY}{\gdef\tsGraphicsScaleY{.9}}{}
\def\checkGraphicsWidth{\ifdim\Gin@nat@width>\linewidth
	\tsGraphicsScaleX\linewidth\else\Gin@nat@width\fi}

\def\checkGraphicsHeight{\ifdim\Gin@nat@height>.9\textheight
	\tsGraphicsScaleY\textheight\else\Gin@nat@height\fi}

\def\fixFloatSize#1{}
\let\ts@includegraphics\includegraphics

\def\inlinegraphic[#1]#2{{\edef\@tempa{#1}\edef\baseline@shift{\ifx\@tempa\@empty0\else#1\fi}\edef\tempZ{\the\numexpr(\numexpr(\baseline@shift*\f@size/100))}\protect\raisebox{\tempZ pt}{\ts@includegraphics{#2}}}}

\AtBeginDocument{\def\includegraphics{\@ifnextchar[{\ts@includegraphics}{\ts@includegraphics[width=\checkGraphicsWidth,height=\checkGraphicsHeight,keepaspectratio]}}}

\def\URL#1#2{\@ifundefined{href}{#2}{\href{#1}{#2}}}

\def\UrlOrds{\do\*\do\-\do\~\do\'\do\"\do\-}%
\g@addto@macro{\UrlBreaks}{\UrlOrds}
\makeatother

\def\fixFloatSize#1{}

\newcolumntype{C}{>{\centering\arraybackslash}X}
  \pgfplotsset{compat=1.14}
\begin{document}
\setcounter{secnumdepth}{3}
\title{Application of quantum scrambling in Rydberg atom on IBM quantum computer}

\author{Daattavya Aggarwal$^{1}$\thanks{E-mail: daggarwal@ph.iitr.ac.in }{ },
              Shivam Raj$^{1}$\thanks{E-mail: shivam.raj@niser.ac.in}{ },
              Bikash K. Behera$^{1}$\thanks{E-mail: bkb13ms061@iiserkol.ac.in} { } \&
              Prasanta K. Panigrahi$^{1}$\thanks{Corresponding author.}\ \thanks{E-mail: pprasanta@iiserkol.ac.in}
              }

\maketitle 

\begin{affiliations}
 \item
      Department of Physics\unskip,  
    Indian Institute of Technology\unskip,  Roorkee\unskip, 247667\unskip, India
    \item
     Department of Physical Sciences\unskip,
    National Institute of Science Education and Research\unskip,  HBNI\unskip, Jatni\unskip, 752050\unskip, Odisha\unskip, India
    \item
      Department of Physical Sciences\unskip, 
    Indian Institute of Science Education and Research Kolkata\unskip, Mohanpur\unskip, 741246\unskip, West Bengal\unskip, India
\end{affiliations}
        
\begin{abstract}

         Quantum scrambling measured by out-of-time-ordered correlator (OTOC) has an important role in understanding the physics of black holes and evaluating quantum chaos. It is known that Rydberg atom has been a general interest due to its extremely favourable properties for building a quantum simulator. Fast and efficient quantum simulators can be developed by studying quantum scrambling in related systems. Here we present a general quantum circuit to theoretically implement an interferometric protocol which is a technique proposed to measure OTOC functions. We apply this circuit to measure OTOC and hence the quantum scrambling in a simulation of a 1-D Ising spin model for Rydberg atom. We apply this method to both initial product and entangled states to compare the scrambling of quantum information in both cases. Finally we discuss other constructions where this technique can be applied.

\end{abstract}


\section{Introduction}

Measuring the magnitude of quantum chaos is an important fundamental problem which has several applications \cite{qsr_FrischNature2000,qsr_PokharelSciRep2018,qsr_ChenarXiv2018}. It is well known that certain quantum mechanical systems are dynamically equivalent to black holes in quantum gravity \cite{qsr_CameliaNature1998,qsr_GiddingsPRD2013,qsr_NomuraPRD2013} which are important theoretical objects studied by physicists \cite{qsr_AharonyPhysRep2000,qsr_BanksPRD1997}. Quantum chaos also has important implications in the fields of quantum information processing and many body quantum systems \cite{qsr_UllmoRPP2008,qsr_KosarXiv2017,qsr_AlessioAP2016}. The scrambling of quantum information has been identified as a good diagnostic tool to measure the magnitude of chaos in a system. Scrambling \cite{qsr_HosurJHEP2016,qsr_SekinoJHEP2008} is a process where a local perturbation spreads over the degrees of freedom of a quantum many body system.
   
Scrambling implies the delocalization of quantum information and once a system has reached a scrambled state, it is impossible to learn about initial perturbations by performing any measurement on the final state. Initially commuting operators grow under time evolution to have large commutators with each other and other operators. Scrambling time refers to the time when the state of the system becomes scrambled. This is a quantum mechanical interpretation of the butterfly effect \cite{qsr_ShenkerJHEP2014}. An object known as the out-of-time-ordered correlator (OTOC) \cite{qsr_LiPRX2017,qsr_MaldacenaJHEP2016,qsr_KitaevFPPS2014} has been proposed to measure the scrambling time. In this work, we present a general theoretical protocol to measure the decay of the OTOC and subsequently measure the scrambling of information in an Ising spin model simulated for Rydberg atoms. While this protocol has been attempted experimentally, it has been performed under certain assumptions. We present a general method to simulate this protocol theoretically which could lead to further probing in the field. It can be applied on strongly correlated quantum many body systems, which have been difficult to study experimentally \cite{qsr_BlochPSSB2010}. We attempt to test the protocol by using the IBM quantum processor, `ibmqx4' by simulating two spin Ising model for Rydberg atoms \cite{qsr_KimIEEE2017,qsr_CortinasIEEE2017,qsr_SchaussQST2018}. The IBM Quantum Experience allows the quantum computing community to access multiple quantum processors and a large number of experiments have been done using their platform \cite{qsr_SrinivasanarXiv2018,qsr_GangopadhyayQIP2018,qsr_AlsinaPRA2016,qsr_GhoshQIP2018,qsr_BeheraarXiv542017,qsr_DasharXiv2017,qsr_GurnaniarXiv2017,qsr_SatyajitarXiv2017,qsr_KandalaNat2017,qsr_VishnuPKarXiv2017,qsr_SisodiaQIP2017,qsr_KalraarXiv2017,qsr_RoyarXiv2017,qsr_ViyuelaNQI2018,qsr_BeheraQIP3122017,qsr_HegadearXiv2017,qsr_DasharXiv782018,qsr_SrinivasanarXiv10928,qsr_Beheraarxiv06530}. As Ising spin model is a two spin level model it is conducive to study through quantum computation techniques by using qubits. 

The scrambling time is measured through the decay of the out-of-time-ordered correlation (OTOC) function.
\begin{equation}
    F(t) = \big \langle W_t^\dagger V^\dagger W_t V \big \rangle
\end{equation}

where, V and W are unitary operators which commute at time $t = 0$. $W_t$ and $U(t)$ are the Heisenberg and time evolution operators defined as, $W_t = U(-t)WU(t)$) and $U(t) = e^{-i\ham  t}$ respectively. Here, $\ham$ represents the time-independent dynamic Hamiltonian of the system. F(t) enables us to measure the time when initially commuting operators, V and W, fail to commute. The relation between the OTOC function and the unitary operators is expressed in the following relation.

\begin{eqnarray}
    & \big \langle \big |\big[ W_t,V \big] \big | ^2 \big \rangle = 2(1 - Re[F])
\end{eqnarray}

Let's consider a system with N spins. For the system to reach a scrambled state, the information in one spin must spread to all the spins \cite{qsr_SwinglePRA2016}. The smallest perturbation in the system involves at least two spins. If unitary operations were to be performed on any two spins at a regular time interval $\delta t$, the scrambling time will be 
\begin{equation}
    t^* = \delta t \log_2 N
\end{equation}
$\big \langle \big | \big[ W_t,V \big] \big | ^2 \big \rangle$ has an order of magnitude of unity at timescales around the scrambling time. Hence Re[F] is an appropriate quantity for assessing scrambling.

\section{Results} 

\textbf{Interferometric Protocol}: It has been shown that F(t) can be measured via many-body interferometry \cite{qsr_ViyuelaNQI2018,qsr_MullerPRL2009,qsr_AbaninPRL2012,qsr_PedernalesPRL2014,qsr_BohrdtNJP2017}.
The interferometric protocol has been discussed by Swingle \emph{et al.} \cite{qsr_SwinglePRA2016,qsr_SwinglearXiv2018}. It lays out a method to measure the OTOC function. However in the experimental realizations of this protocol, some assumptions have always been made owing to the extreme challenges of the experimental setup. The interferometric protocol involves backward and forward evolution in time under an identical Hamiltonian. Realizing this in practice without dissipative effects is a difficult challenge. We present a quantum circuit to theoretically simulate this protocol. As we are considering a simulation, this avoids the inherent experimental challenges \cite{qsr_SwinglePRA2016} and allows us to measure the OTOC function without considering the dissipative effects. The interferometric protocol has been described by Swingle et.al. Here we present a generic quantum circuit to implement it \cite{qsr_SwinglePRA2016,qsr_SwinglearXiv2018}.

\par
\begin{figure}[H]
\includegraphics[scale=0.95]{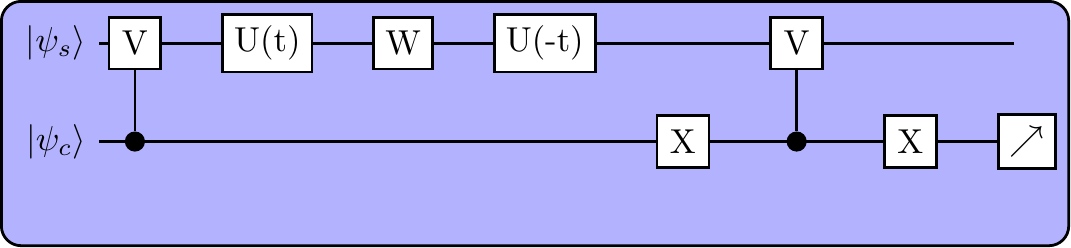}
\caption{\textbf{Circuit implementing the general interferometric protocol}. Here $\ket{\psi_s}$ and $\ket{\psi_c}$ represent the states of the system and the control qubit respectively. V and W are the initially commuting unitary operators. U(t) is the time-evolution operator. Controlled-V, W and U(t) are decomposed into a product of single qubit and CNOT gates to design the quantum circuit in the IBM quantum processor, ibmqx4. The measurement of the control qubit is performed in the X-basis.}
\label{qsr_fig1}
\end{figure}

 Consider a n-qubit quantum system initially in the state ($| \psi \rangle _s$) and a control qubit ($\ket{\psi_c}$) initially in the state $\ket{0}$. By using the control qubit, we can easily prepare multiple branches for the system by applying appropriate controlled operations. We apply a Hadamard gate to transform the control qubit into the state $\frac{|0 \rangle + |1 \rangle}{\sqrt{2}}$. Then we prepare a final state where one branch undergoes the operation $V W_t$ and the other branch undergoes the operation $W_t V$. The OTOC function F(t), measures the overlap between these two states. This is achieved by preparing the following resultant state,
\begin{equation}
    | \psi _f \rangle = \frac{(V W_t |\psi\rangle_s ) |0\rangle_c + (W_t V |\psi\rangle _s)|1 \rangle _c}{\sqrt{2}}
\end{equation}
The control qubit is then measured in the X-basis and the expectation value of the control qubit, $\big \langle X \big \rangle$ is equal to Re[F]. 
This can be obtained via the generic quantum circuit as depicted in Fig. \ref{qsr_fig1}.

Coupling of neutral atoms to highly excited Rydberg states has been shown to be a promising way to simulate the Ising spin model \cite{qsr_KimIEEE2017,qsr_SchaussQST2018,qsr_LabuhnNature2016,qsr_NguyenPRX2018,qsr_AugerPRA2018}. Consider a driving  resonant laser with Rabi frequency $\Omega$ that couples atoms in a ground state $\ket{g_i}$ to a highly excited Rydberg state $\ket{r_i}$. The resultant Rydberg atom pairs face strong, repulsive van der Waals interactions of the type $V_{ij} = \frac{C}{R_{ij}}$, where $R_{ij}$ is the distance between Rydberg atom pairs (i, j) and C $\textgreater$ 0. The general Hamiltonian for an Ising spin model for Rydberg atoms \cite{qsr_KimIEEE2017} can be represented by 
\begin{equation}
    \ham = \hbar \Omega \sum_{i} \hat{\sigma_x}^i + \sum_{i<j} V_{ij} \hat{n_i}\hat{n_j}          ,
\end{equation}
where $\hbar$ is the reduced Planck's constant, $\hat{\sigma_x}^i$ is the Pauli-X operator acting on the $i^{th}$ spin and $\hat{n}$ is Rydberg atom number.
The total spin operator is given as
\begin{equation}
S_Z = \sum_{i} \hat{\sigma_z}^i
\end{equation}
where $\hat{\sigma_z}^i$ is the Pauli-Z operator acting on the $i^{th}$ spin.
The unitary operators V and W are chosen such that they commute at time t = 0 and satisfy the required conditions for the application of the interferometric protocol. V and W are defined to be, $V = W = e^{-i \phi S_Z}$, where $\phi = \pi/4$.  The time-evolution operator U(t) and the commuting operators V and W are decomposed into unitary gates which can be directly applied on qubits.
Without loss of generality, we can assume $\ket{0}$ and $\ket{1}$ to represent the ground state and the excited state of the Rydberg atom respectively. This allows us to perform the simulation using a quantum computer. We consider two different initial states to illustrate the power of this technique. For the first set of data, we consider an initial state where both the qubits are in the ground state, i.e., $\ket{\psi _s} = \ket{00}$. In this case, the interactions between the qubits develop as a result of the dynamics of the system and are expressed through the time evolution operator U(t). For the second set of data, we consider an initial state where the qubits are in an entangled state. The qubits form Rydberg atom pairs under the condition of Rydberg blockades \cite{qsr_UrbanNatPhys2009,qsr_GaetanNatPhys2009,qsr_WilkPRL2010,qsr_ZengPRL2017,qsr_ZhaoNature2017} which implies that both the qubits cannot be simultaneously excited to Rydberg states. This is done by first preparing both the qubits in $\ket{00}$ state. Then a Pauli-X gate and a Hadamard gate are applied on the first and second qubit respectively. A CNOT gate is applied taking the second qubit as the control qubit and the first qubit as the target qubit. After the operation of the above gates, the resultant state is found to be, $\ket{\psi_s} = \frac{\ket{01} + \ket{10}}{\sqrt{2}}$. Now we can easily see that both of the qubits are not in the excited state simultaneously which confirms that they act as a Rydberg blockade. Owing to the entanglement of the system, we expect to have different results from the first case. The interaction in this system is not solely due to time evolution but is also present due to entanglement correlations. To showcase the general technique as described above, we simulate the above model using IBM quantum experience platform.

\textbf{Observations}: 
\begin{figure}[H]
\includegraphics[scale = 0.40]{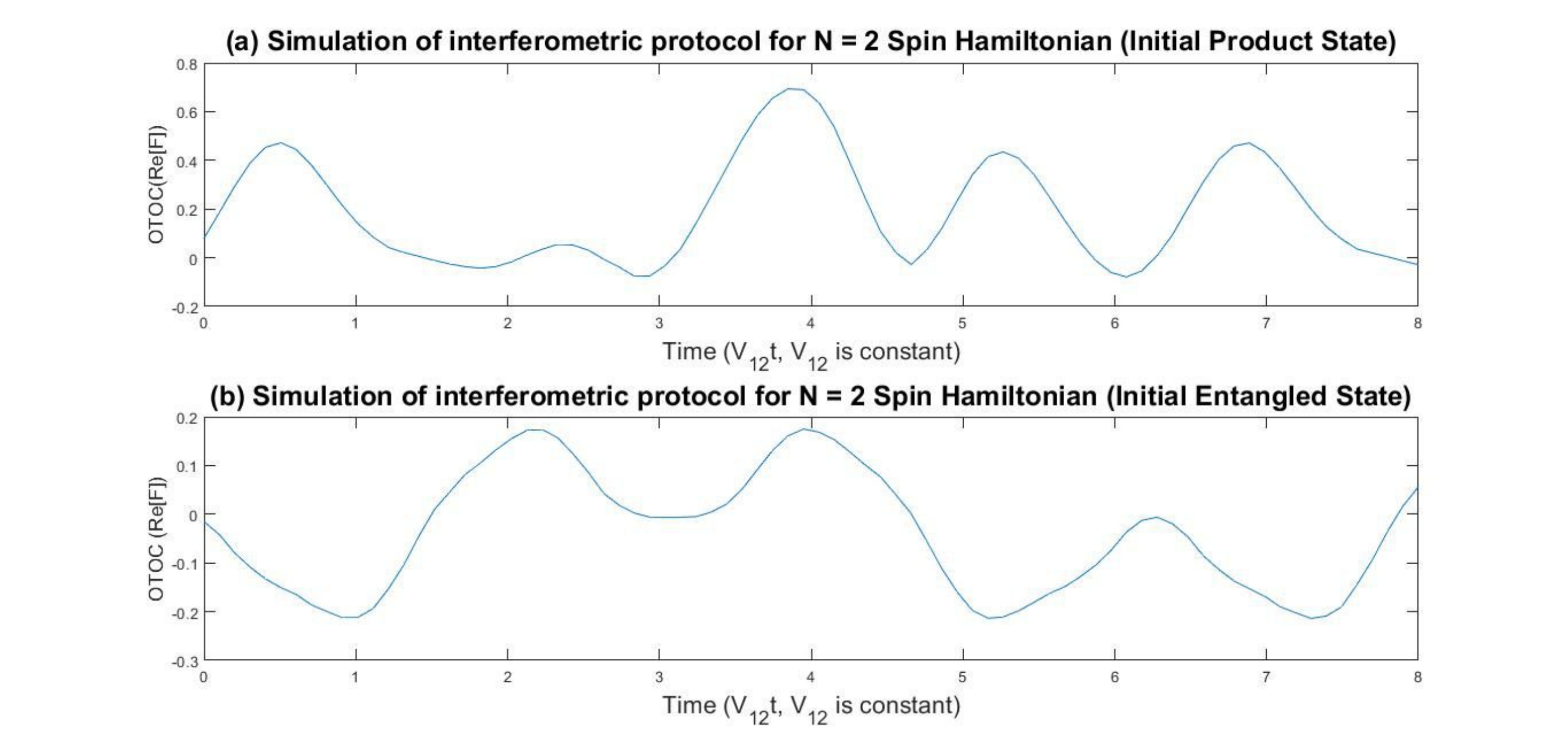}
\caption{\textbf{out-of-time-ordered correlator (OTOC) function behaviour (Simulation Result) }: \textbf{(a)} OTOC (Re[F]) as a function of time for a two spin (N = 2) Ising model Rydberg atom when the initial state is $|\psi_s\rangle = |00\rangle$. It is clearly seen that the initial state of both the Rydberg atoms is in the ground state. The graph is obtained by simulating the quantum circuit on the `ibmqx4' quantum processor. It is to be noted that the data has not been taken on the real quantum computer and hence dissipation and noise can be ignored.  As we can observe, the OTOC after an initial rise decays from $V_{12}t = 1$ to $V_{12}t = 3$. After this point of time, it undergoes oscillations of large magnitude. The OTOC seems to follow a periodic pattern. From the graph, it is observed that quantum information travels periodically from one end of the system to the other. It does not reach an equilibrium value for the time period of the experiment ($V_{12} t$ = 0 to $V_{12} t$ = 8). \textbf{(b)} OTOC (Re[F]) as a function of time for a two spin (N = 2) Ising model for Rydberg atom when the initial state is $|\psi_s\rangle = \frac{\ket{01} + \ket{10}}{\sqrt{2}}$. It is clearly seen that the qubits form Rydberg atom pairs and exist in an entangled state. The graph is plotted after simulating on the `ibmqx4' quantum processor.  As we can notice, the OTOC does not follow any regular pattern. In this case quantum information travels from one end of the system to the other as well. As a whole, the magnitude of the OTOC is much lesser than the case when the initial state is a product state (\textbf{Case (a)}) which can have important implications in the future.}
\label{qsr_fig2}
\end{figure}

\begin{figure}[H]
\includegraphics[scale=0.44]{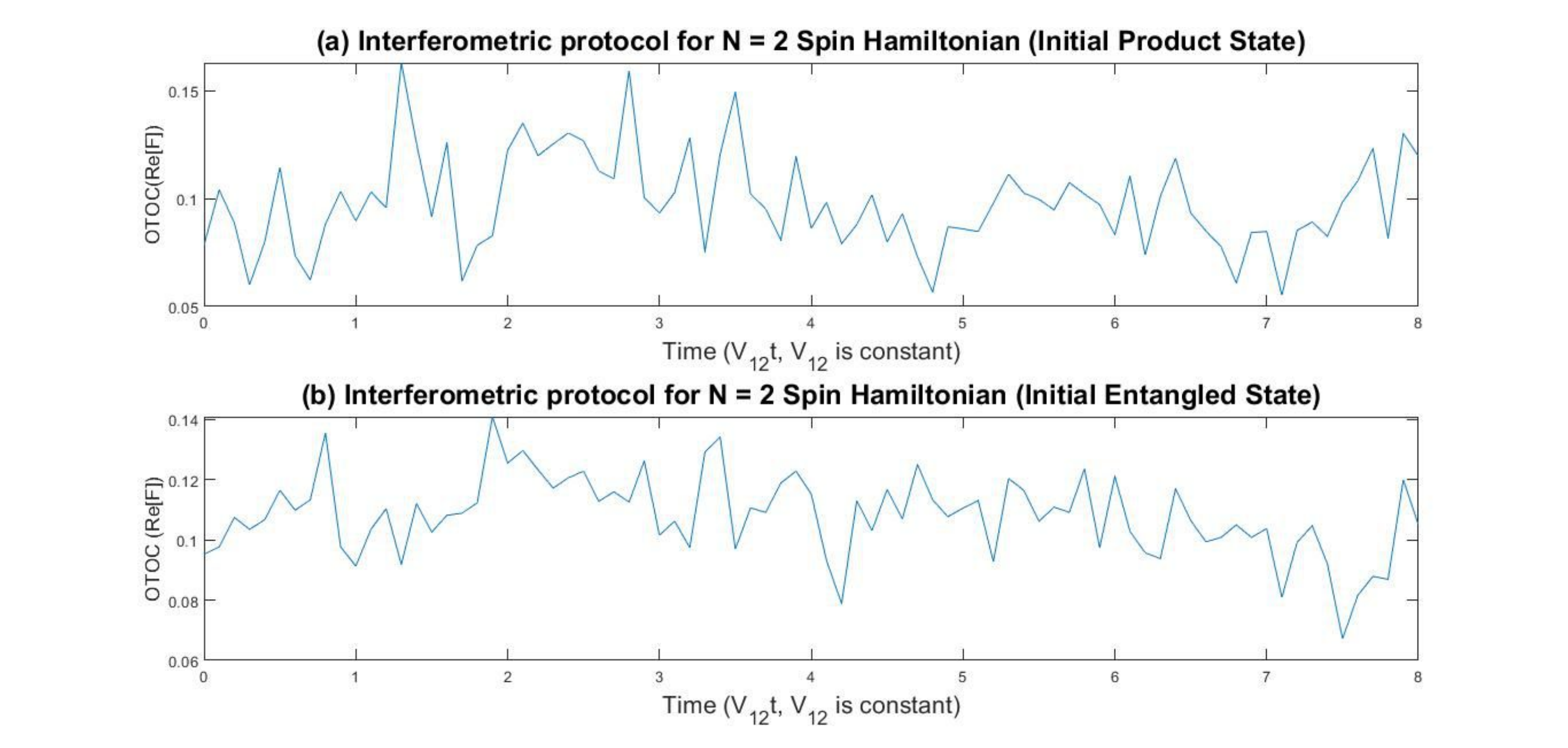}
\caption{\textbf{out-of-time-ordered correlator (OTOC) function behaviour (Experimental Result)}: \textbf{(a)} OTOC (Re[F]) as a function of time for a two spin (N = 2) Ising model Rydberg atom when the initial state is $|\psi_s\rangle = |00\rangle$. It is observed that the initial state of both the Rydberg atoms is the ground state. The graph is obtained by running the quantum circuit on the `ibmqx4' quantum processor, i.e. the results are collected after performing the experiment on the real quantum chip. In this case, the effects of dissipation and noise can not be ignored and are clearly visible. It can be observed, the OTOC does not follow a regular pattern. It does not reach an equilibrium value for the time period of the experiment ($V_{12} t$ = 0 to $V_{12} t$ = 8). \textbf{(b)} OTOC (Re[F]) as a function of time for a two spin (N = 2) Ising model for Rydberg atom when the initial state is $|\psi_s\rangle = \frac{\ket{01} + \ket{10}}{\sqrt{2}}$. It is clearly seen that the qubits form Rydberg atom pairs and exist in an entangled state. The graph is obtained by running the quantum circuit on the `ibmqx4' quantum processor, i.e. the results are compiled after execution of the quantum circuit on the real quantum chip.  As we can notice, the OTOC does not follow any regular pattern. In the graph, quantum information travels from one end of the system to the other. As a whole, the magnitude of the OTOC is \textit{larger} than the case when the initial state is a product state. This is in direct contrast to the simulation results. From these results it is evident that environmental noise and dissipation greatly affect OTOC measurements.}
\label{qsr_fig3}
\end{figure}

\section{Discussion}
These results shed light on scrambling of quantum information in the chosen system. The results have been collected through simulation on the `ibmqx4' quantum processor. two system qubits and one control qubit have been used. In Fig. \ref{qsr_fig2} (a), the initial state was taken as $\ket{00}$ and the qubits were not entangled. So the chosen system represents the simplest case of the chosen model where both the qubits are in the ground state. It has been recently observed that the OTOC decays quickly in large many-body systems \cite{qsr_SwinglePRA2016}. In our case, despite having a relatively smaller system, a considerable decay in the OTOC is observed. In Fig. \ref{qsr_fig2} (b), the initial state is taken as $\frac{\ket{01} + \ket{10}}{\sqrt{2}}$ which is an entangled state. It is observed that, the magnitude of the OTOC is considerably smaller than the magnitude observed in Fig. \ref{qsr_fig2} (a). However in this case, no pattern in the OTOC is recognized. The decay is also not well defined. In Fig. \ref{qsr_fig3}, it is difficult to draw conclusions due to the prevalence of environmental noise. To combat this challenge, a renormalization procedure has been proposed by Swingle and Halpern \cite{qsr_SwinglearXiv2018}. The renormalization procedure allows scrambling measurements to be resilient to environmental noise. This procedure can be easily extended to measure scrambling of quantum information in much larger systems with varying degrees of interaction \cite{qsr_BernienNature2017}. Modifying the circuit for larger qubit systems, although non-trivial, can be done by appropriately simulating the Hamiltonian and operators while keeping in mind the universality of quantum gates.   

\section{Outlook}
As more sophisticated and larger quantum processors are developed, it will be possible to use this technique to measure more complicated systems \cite{qsr_SwinglePRA2016,qsr_BernienNature2017}. We feel that important fundamental breakthroughs can be made in the field of quantum chaos \cite{qsr_ChenarXiv2018,qsr_IyodaPRA2018} by simulating classically chaotic models using techniques similar to the one presented in this paper. Also further probing might reveal important links between faster computation and quantum scrambling and the dissemination of quantum information \cite{qsr_BrownPRL2016}. Developing a quantum system with interactions similar to those in a black hole is an ongoing challenge. If such a strongly correlated, quantum mechanical model was to be developed, we can measure the scrambling of information which can lead to verification of the fast scrambling hypothesis \cite{qsr_SekinoJHEP2008,qsr_LashkariJHEP2013}. Links between the hiding of quantum information and quantum chaos can also be elucidated \cite{qsr_HosurJHEP2016}.

\section{Methods}
\textbf{Quantum Circuit for simulating operators V and W}
The unitary operator V is defined as,
\begin{equation}
    V = e^{ -i \frac{\pi}{4} S_Z }
\end{equation}
where $S_Z$ is calculated as, 
\begin{equation}
     S_Z =  \hat{\sigma_z}^1 \otimes I + I \otimes\hat{\sigma_z}^2
\end{equation}
 
V may be decomposed into a product of two-level unitary matrices, which act non-trivially on vectors in the orthonormal basis. Now our immediate goal is to construct a circuit implementing V. This circuit can be built using only single qubit and CNOT gates owing to the universality of quantum computation. To achieve this, we construct a Gray code sequence which links the two-level unitary matrices to ultimately implement V. A Gray code is a sequence of binary numbers which links an initial state to a final state such that any two neighbouring numbers differ only in a single bit. With the help of Gray codes we rotate a multi-qubit system into a state where the corresponding non-trivial, single qubit unitary transformation can be applied directly \cite{qsr_sup_Nielsen2002,qsr_sup_LiarXiv2012}.
\begin{figure}[H]
\includegraphics[scale = 1.0]{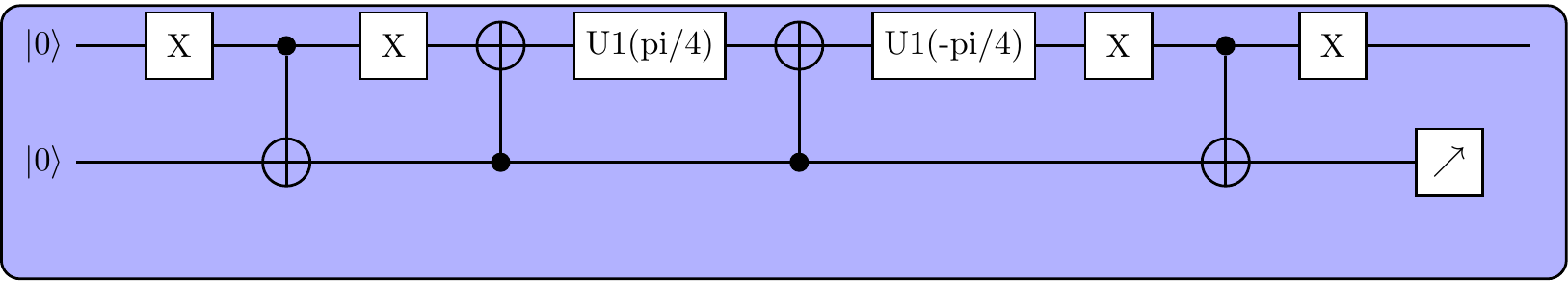}
\caption{Here $U_1$ is a physical gate provided by IBM quantum experience which represents a general rotation on the Bloch sphere and has one controllable parameter. This is the circuit of the unitary operator V after being decomposed into a product of two level unitary matrices. The circuit is designed with the help of CNOT gates and single qubit gates. The CNOT operation means when the control qubit is 1, then the target qubit is flipped. If the control qubit is 0, then the target qubit remains unchanged. The control qubit remains unchanged in both the cases.} 
\label{qsr_sup_Fig1}
\end{figure}
   
\textbf{Simulation of Hamiltonian}
For simulation of the Hamiltonian we employ a first order Trotter decomposition \cite{qsr_sup_Nielsen2002,qsr_sup_HegadearXiv2017}.
\begin{equation*}
    e^{-i\ham t} = e^{-i \ham_1 t}e^{-i \ham_2 t}...e^{-i \ham_n t} +O\big({t^2}\big)
\end{equation*}
where $\ham_1, \ham_2,...,\ham_n$ are Hamiltonians acting on local subsystems involving k-qubits of an n-qubit system. The system Hamiltonian is, $\ham = \sum_{1} ^ {n} \ham_k$. The Hamiltonian is then decomposed into a sequence of unitary transformations which can be implemented through any set of universal quantum gates. In the model we have chosen above, the Hamiltonian is
\begin{equation*}
    \ham = \hbar\Omega\hat{\sigma_x}^1 + \hbar\Omega\hat{\sigma_x}^1 + V_{12}\hat{\sigma_z}^1\hat{\sigma_z}^2
\end{equation*}
To implement the Trotter decomposition, we use $\ham = \ham_1 + \ham_2 + \ham_3$. Without loss of generality, we can assume a system of units such that $\hbar\Omega = 1$ to simplify the calculations.
\begin{equation*}
    \ham_1 = \hat{\sigma_x} ^1 \otimes I
\end{equation*}
After exponentiation, we finally get
$
\begin{bmatrix}
    \cos{t} & -\iota\sin{t}\\
    -\iota\sin{t} & \cos{t} 
\end{bmatrix}
$ acting on the first qubit and identity matrix acting on the second. Here t is the time elapsed since the beginning of the experiment. As the IBM Q Experience is a static system, by taking t as a controllable parameter we are able to effectively simulate the interferometric protocol. $H_2$ and $H_3$ can be implemented in a similar way \cite{qsr_sup_Whitfield2010}. For producing the net Hamiltonian, we simply have to multiply $e^{-i\ham_1t}$, $e^{-i\ham_2t}$ and $e^{-i\ham_3t}$. The corresponding circuit is given as follows.

\begin{figure}[H]
    \includegraphics[scale = 0.83]{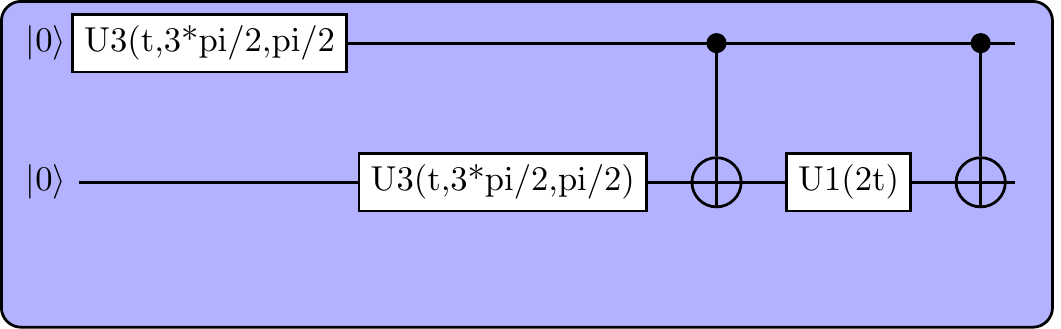}
    \caption{Here $U_3$ is a physical gate provided by IBM Q experience which represents a general rotation on the Bloch sphere and has 3 controllable parameters. As the IBM Q experience is a static system and we needed to study the dynamics of the system, we took the time `t' as a controllable parameter. The values $\frac{3\pi}{2}$ and $\frac{\pi}{2}$ come from the appropriate decomposition of the U(t) transformation into two-level unitary matrices. $U_1$ is also a general physical gate provided by IBM Q experience and has one controllable parameter. The values of those parameters are also shown here.}
\label{qsr_sup_Fig2}  
\end{figure}

\begin{figure}[H]
\centering
\includegraphics[scale=0.4]{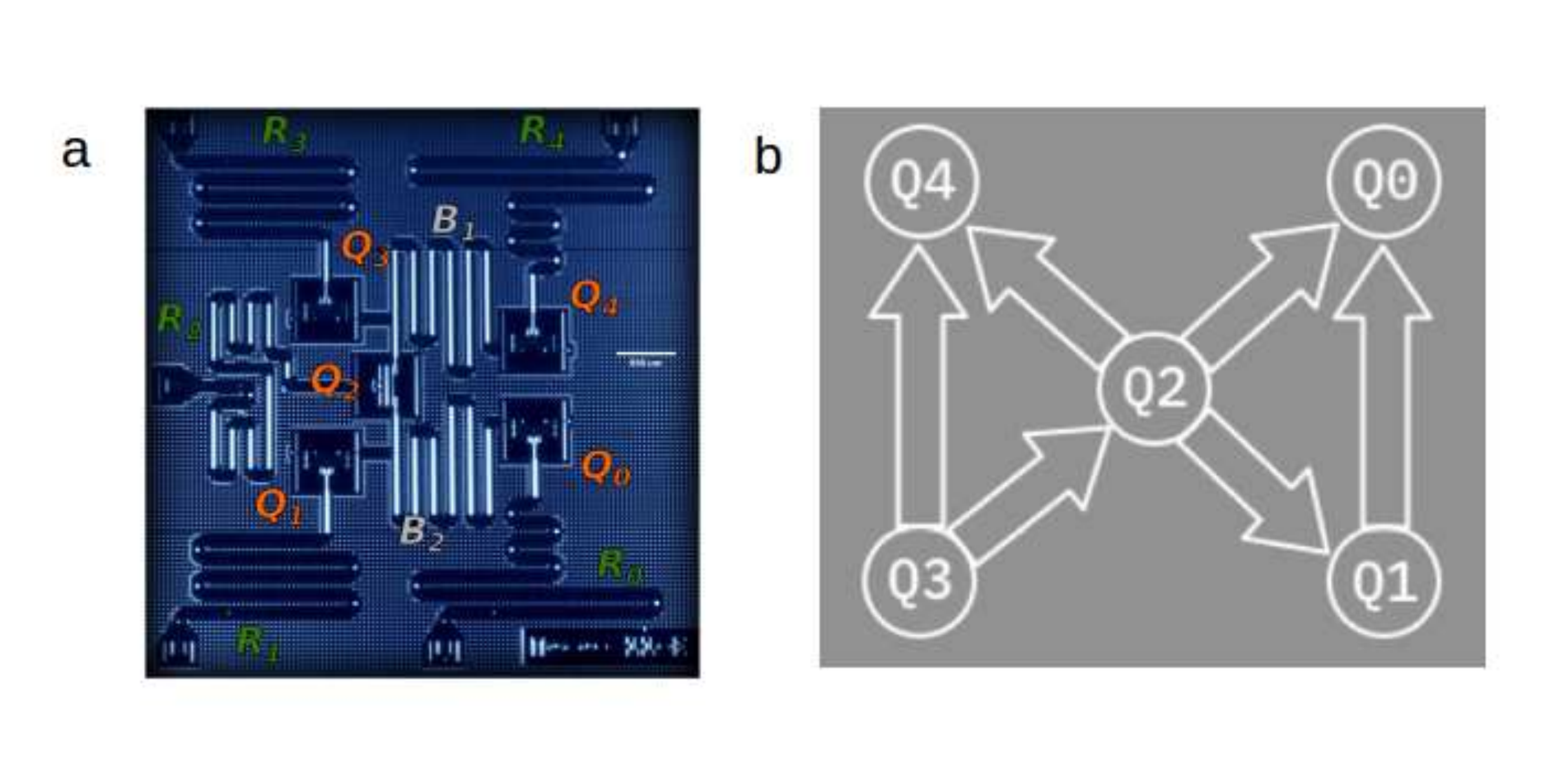}
\caption{\textbf{(a)} A schematic diagram of the chip layout of 5-qubit quantum processor `ibmqx4'. The chip is maintained in a dilution refrigerator at temperature 0.021 K. All 5 transmon qubits are connected with the two coplanar waveguide (CPW) resonators as shown. q[2], q[3], q[4] and q[0], q[1], q[2] are coupled with the two CPWs with resonance frequencies 6.6 GHz and 7.0 GHz respectively. Individual qubits in the chip are controlled and readout by particular CPWs. \textbf{(b)} The CNOTs coupling map is as follows : $\{q1 \rightarrow (q[0]), q2 \rightarrow (q[0], q[1], q[4]), q[3] \rightarrow (q[2], q[4])\}$, where $i \rightarrow (j)$ means $i$ and $j$ denote the control and the target qubit respectively for implementing CNOT gate. The errors in gates and readout are of the order $10^{-2}$ to $10^{-3}$.}
\label{qsr_sup_Fig6}
\end{figure}

\textbf{Experimental Architecture}
The experimental device parameters of `ibmqx4' chip are listed in Table \ref{qsr_sup_tab1}. The readout resonator's resonance frequency, qubit frequency, anharmonicity, qubit-cavity coupling strength, relaxation time and coherence time are respectively denoted by $\omega^{R}_{i}$, $\omega_{i}$, $\delta_{i}$, $\chi$, $T_1$ and $T_2$. The connection and control of five superconducting qubits (q[0], q[1], q[2], q[3] and q[4]) is shown in Fig. \ref{qsr_sup_Fig6} \textbf{(b)}. The single-qubit and two-qubit controls are provided by the coplanar wave guide (CPW) resonators. The black and white lines denote the control respectively. The qubits q[2], q[3], q[4] and q[0], q[1], q[2] are coupled via two superconducting CPWs, with resonator frequencies of 6.6 GHz and 7.0 GHz respectively. All the qubits are controlled and read out by different CPWs. The quantum chip, `ibmqx4' is cooled in a dilution refrigerator at a temperature of 0.021 K. The single-qubit gate error is of the order $10^{-3}$. The multi-qubit and readout error are of the order $10^{-2}$. Randomized benchmarking is used to measure the gate errors.

\textbf{Data availability.} The data that support the findings of this study are available from the corresponding author upon reasonable request.

\begin{table}[H]
\centering
\begin{tabular}{ c c c c c c }
\hline
\hline
Qubits & $\omega^{\star}$ (GHz) & $T^{||}_{1}$ ($\mu s$) & $T^{\perp}_{2}$ ($\mu s$) & GE$^{\dagger}$ & RE$^{\ddagger}$ \\
\hline
q[0] & 5.24 & 48.80 & 14.70 & 0.86 & 7.00 \\
q[1] & 5.31 & 49.60 & 55.00 & 1.29 & 5.80 \\
q[2] & 5.35 & 48.00 & 32.60 & 1.20 & 8.60 \\ 
q[3] & 5.41 & 35.60 & 23.60 & 3.78 & 3.70 \\
q[4] & 5.19 & 55.20 & 31.90 & 1.03 & 5.80 \\
\hline
\hline
\end{tabular}\\
$\star$ Frequency, $||$ Relaxation time, $\perp$ Coherence time, $\dagger$ Gate Error, $\ddagger$ Readout Error \\
\caption{\textbf{Experimental parameters of the device `ibmqx4' are presented.}}
\label{qsr_sup_tab1}
\end{table}

\bibliographystyle{naturemag}

\section*{Acknowledgements} The authors acknowledge the suggestions and comments provided by Nicole Yunger Halpern (Caltech). D.A. acknowledges the hospitality of IISER Kolkata during the project work. S.R. and B.K.B. acknowledge financial support of Inspire fellowship provided by Department of Science and Technology (DST), Govt. of India. We acknowledge the support of IBM Quantum Experience for providing access to the IBM quantum processors. The views expressed are those of the authors and do not reflect the official position of IBM or the IBM Quantum Experience team. 

\section*{Author contributions}
Theoretical analysis, design of quantum circuit and simulation was performed by D.A. Collection and analysis of data was done by D.A. and S.R. The project was supervised by B.K.B. A thorough checking of the manuscript was done by P.K.P. D.A., S.R. and B.K.B. have completed the project under the guidance of P.K.P.   

\section*{Competing interests}
The authors declare no competing financial interests. Readers are welcome to comment on the online version of the paper. Correspondence and requests for materials should be addressed to P.K.P. (pprasanta@iiserkol.ac.in).
\section{Supplementary Information: Application of quantum scrambling in Rydberg atom on IBM quantum computer}
For simulating the interferometric protocol, we used QISKit to take both simulation and experimental results. The qasm code for the same is as follows: 

\begin{lstlisting}[language=Python]
{
 "cells": [
  {
   "cell_type": "code",
   "execution_count": null,
   "metadata": {},
   "outputs": [],
   "source": [
    "from qiskit import QuantumProgram\n",
    "import Qconfig\n",
    "import qiskit as qiskit\n",
    "from IBMQuantumExperience import IBMQuantumExperience\n",
    "from math import pi\n"
   ]
  },
  {
   "cell_type": "code",
   "execution_count": null,
   "metadata": {},
   "outputs": [],
   "source": [
    "qp = QuantumProgram()\n",
    "\n",
    "# Creating Registers\n",
    "# create your first Quantum Register called \"qr\" with 2 qubits \n",
    "qr = qp.create_quantum_register('qr', 3)\n",
    "# create your first Classical Register  called \"cr\" with 2 bits\n",
    "cr = qp.create_classical_register('cr', 5)\n",
    "\n",
    "# Creating Circuits\n",
    "# create your first Quantum Circuit called \"qc\" involving your Quantum Register \"qr\"\n",
    "# and your Classical Register \"cr\"\n",
    "qc = qp.create_circuit('Circuit', [qr], [cr])"
   ]
  },
  {
   "cell_type": "code",
   "execution_count": null,
   "metadata": {},
   "outputs": [],
   "source": [
    "Q_SPECS = {\n",
    "    'circuits': [{\n",
    "        'name': 'Circuit',\n",
    "        'quantum_registers': [{\n",
    "            'name': 'qr',\n",
    "            'size': 3\n",
    "        }],\n",
    "        'classical_registers': [{\n",
    "            'name': 'cr',\n",
    "            'size': 5\n",
    "        }]}],\n",
    "}"
   ]
  },
  {
   "cell_type": "code",
   "execution_count": null,
   "metadata": {},
   "outputs": [],
   "source": [
    "qp = QuantumProgram(specs=Q_SPECS)"
   ]
  },
  {
   "cell_type": "code",
   "execution_count": null,
   "metadata": {},
   "outputs": [],
   "source": [
    "# get the circuit by Name\n",
    "circuit = qp.get_circuit('Circuit')\n",
    "\n",
    "# get the Quantum Register by Name\n",
    "quantum_r = qp.get_quantum_register('qr')\n",
    "\n",
    "# get the Classical Register by Name\n",
    "classical_r = qp.get_classical_register('cr')"
   ]
  },
  {
   "cell_type": "code",
   "execution_count": null,
   "metadata": {},
   "outputs": [],
   "source": [
    "\n",
    "circuit.h(quantum_r[0])\n",
    "#####################    Comment out this block if you don't want Initially entangled states  #############\n",
    "circuit.h(quantum_r[1])\n",
    "circuit.x(quantum_r[2])\n",
    "circuit.cx(quantum_r[1],quantum_r[2])\n",
    "#####################    Intially entangled States   ####################\n",
    "circuit.cx(quantum_r[0],quantum_r[1])\n",
    "circuit.ccx(quantum_r[0],quantum_r[1],quantum_r[2])\n",
    "circuit.cx(quantum_r[0],quantum_r[1])\n",
    "circuit.ccx(quantum_r[0],quantum_r[2],quantum_r[1])\n",
    "circuit.cu1(-pi/4,quantum_r[0],quantum_r[1])\n",
    "circuit.ccx(quantum_r[0],quantum_r[2],quantum_r[1])\n",
    "circuit.cu1(pi/4,quantum_r[0],quantum_r[1])\n",
    "circuit.cx(quantum_r[0],quantum_r[1])\n",
    "circuit.ccx(quantum_r[0],quantum_r[1],quantum_r[2])\n",
    "circuit.cx(quantum_r[0],quantum_r[1])\n",
    "\n",
    "\n",
    "#################################### Time Evolution Operator ##############################\n",
    "circuit.u3(4,3*pi/2,pi/2,quantum_r[1])\n",
    "circuit.iden(quantum_r[2])\n",
    "circuit.u3(4,3*pi/2,pi/2,quantum_r[2])\n",
    "circuit.iden(quantum_r[1])\n",
    "circuit.cx(quantum_r[1],quantum_r[2])\n",
    "circuit.u1(8,quantum_r[2])\n",
    "circuit.cx(quantum_r[1],quantum_r[2])\n",
    "####################################         End            ###############################\n",
    "    \n",
    "\n",
    "circuit.x(quantum_r[1])\n",
    "circuit.cx(quantum_r[1],quantum_r[2])\n",
    "circuit.x(quantum_r[1])\n",
    "circuit.cx(quantum_r[2],quantum_r[1])\n",
    "\n",
    "circuit.u1(pi,quantum_r[1])\n",
    "circuit.cx(quantum_r[2],quantum_r[1])\n",
    "circuit.t(quantum_r[1])\n",
    "circuit.x(quantum_r[1])\n",
    "circuit.cx(quantum_r[1],quantum_r[2])\n",
    "circuit.x(quantum_r[1])\n",
    "\n",
    "\n",
    "\n",
    "#################################### Time Evolution Operator ##############################\n",
    "circuit.u3(-4,3*pi/2,pi/2,quantum_r[1])\n",
    "circuit.iden(quantum_r[2])\n",
    "circuit.u3(-4,3*pi/2,pi/2,quantum_r[2])\n",
    "circuit.iden(quantum_r[1])\n",
    "circuit.cx(quantum_r[1],quantum_r[2])\n",
    "circuit.u1(-8,quantum_r[2])\n",
    "circuit.cx(quantum_r[1],quantum_r[2])\n",
    "####################################         End            ###############################\n",
    "\n",
    "\n",
    "circuit.x(quantum_r[0])\n",
    "circuit.cx(quantum_r[1],quantum_r[2])\n",
    "circuit.ccx(quantum_r[0],quantum_r[1],quantum_r[2])\n",
    "circuit.cx(quantum_r[1],quantum_r[2])\n",
    "circuit.ccx(quantum_r[0],quantum_r[2],quantum_r[1])\n",
    "circuit.cu1(-pi/4,quantum_r[0],quantum_r[1])\n",
    "circuit.ccx(quantum_r[0],quantum_r[2],quantum_r[1])\n",
    "circuit.cu1(pi/4,quantum_r[0],quantum_r[1])\n",
    "circuit.cx(quantum_r[0],quantum_r[1])\n",
    "circuit.ccx(quantum_r[0],quantum_r[1],quantum_r[2])\n",
    "circuit.cx(quantum_r[0],quantum_r[1])\n",
    "circuit.x(quantum_r[0])\n",
    "circuit.h(quantum_r[0])\n",
    "circuit.measure(quantum_r[0], classical_r[0])\n",
    "\n"
   ]
  },
  {
   "cell_type": "code",
   "execution_count": null,
   "metadata": {},
   "outputs": [],
   "source": [
    "qp.get_circuit_names()"
   ]
  },
  {
   "cell_type": "code",
   "execution_count": null,
   "metadata": {},
   "outputs": [],
   "source": [
    "QASM_source = qp.get_qasm('Circuit')\n",
    "\n",
    "print(QASM_source)"
   ]
  },
  {
   "cell_type": "code",
   "execution_count": null,
   "metadata": {},
   "outputs": [],
   "source": []
  },
  {
   "cell_type": "code",
   "execution_count": null,
   "metadata": {
    "scrolled": true
   },
   "outputs": [],
   "source": []
  },
  {
   "cell_type": "code",
   "execution_count": null,
   "metadata": {},
   "outputs": [],
   "source": [
    "backend = 'local_qasm_simulator' \n",
    "circuits = ['Circuit']  # Group of circuits to execute"
   ]
  },
  {
   "cell_type": "code",
   "execution_count": null,
   "metadata": {},
   "outputs": [],
   "source": [
    "qobj=qp.compile(circuits, backend,shots=8192) # Compile your program"
   ]
  },
  {
   "cell_type": "code",
   "execution_count": null,
   "metadata": {},
   "outputs": [],
   "source": [
    "\n",
    "result = qp.run(qobj, timeout=240)\n",
    "print(result)"
   ]
  },
  {
   "cell_type": "code",
   "execution_count": null,
   "metadata": {},
   "outputs": [],
   "source": [
    "result.get_counts('Circuit')"
   ]
  },
  {
   "cell_type": "code",
   "execution_count": null,
   "metadata": {},
   "outputs": [],
   "source": [
    "ran_qasm = result.get_ran_qasm('Circuit')\n",
    "\n",
    "print(ran_qasm)"
   ]
  },
  {
   "cell_type": "code",
   "execution_count": null,
   "metadata": {},
   "outputs": [],
   "source": [
    "\n",
    "out = qp.execute(circuits, backend, timeout=240)\n",
    "print(out)"
   ]
  },
  {
   "cell_type": "code",
   "execution_count": null,
   "metadata": {
    "scrolled": false
   },
   "outputs": [],
   "source": [
    "qp.set_api(Qconfig.APItoken, Qconfig.config['url']) # set the APIToken and API url\n",
    "real_device_backend = [backend for backend in qp.online_devices() if qp.get_backend_configuration(backend)['n_qubits'] == 5 and qp.get_backend_status(backend)['available'] == True]\n",
    "# find an appropriate real device backend that your APIToken has access to run that has 5 qubits and is available"
   ]
  },
  {
   "cell_type": "code",
   "execution_count": null,
   "metadata": {},
   "outputs": [],
   "source": [
    "backend = real_device_backend[0]  # Backend where you execute your program; in this case, on the real Quantum Chip online \n",
    "\n",
    "circuits = ['Circuit']   # Group of circuits to execute\n",
    "shots = 8192           # Number of shots to run the program (experiment); maximum is 8192 shots.\n",
    "max_credits = 5          # Maximum number of credits to spend on executions. \n",
    "\n",
    "result_real = qp.execute(circuits, backend=backend, shots=shots, max_credits=max_credits, timeout=50)\n"
   ]
  },
  {
   "cell_type": "code",
   "execution_count": null,
   "metadata": {},
   "outputs": [],
   "source": [
    "result_real.get_counts('Circuit')"
   ]
  },
  {
   "cell_type": "code",
   "execution_count": null,
   "metadata": {},
   "outputs": [],
   "source": [
    "backend = real_device_backend[0]\n",
    "backend"
   ]
  },
  {
   "cell_type": "code",
   "execution_count": null,
   "metadata": {},
   "outputs": [],
   "source": [
    "my_jobs = qp.get_api().get_jobs(limit=100)\n",
    "done_jobs = [j for j in my_jobs if j['status']=='COMPLETED' ]\n",
    "for j in done_jobs:\n",
    "    for q in j['qasms']:\n",
    "        print(q['qasm'])\n",
    "        print(q['result'])\n",
    "        "
   ]
  },
  {
   "cell_type": "code",
   "execution_count": null,
   "metadata": {},
   "outputs": [],
   "source": []
  },
  {
   "cell_type": "code",
   "execution_count": null,
   "metadata": {},
   "outputs": [],
   "source": []
  },
  {
   "cell_type": "code",
   "execution_count": null,
   "metadata": {},
   "outputs": [],
   "source": []
  },
  {
   "cell_type": "code",
   "execution_count": null,
   "metadata": {},
   "outputs": [],
   "source": []
  },
  {
   "cell_type": "code",
   "execution_count": null,
   "metadata": {},
   "outputs": [],
   "source": []
  }
 ],
 "metadata": {
  "kernelspec": {
   "display_name": "Python 3",
   "language": "python",
   "name": "python3"
  },
  "language_info": {
   "codemirror_mode": {
    "name": "ipython",
    "version": 3
   },
   "file_extension": ".py",
   "mimetype": "text/x-python",
   "name": "python",
   "nbconvert_exporter": "python",
   "pygments_lexer": "ipython3",
   "version": "3.6.4"
  }
 },
 "nbformat": 4,
 "nbformat_minor": 2
}
\end{lstlisting}

\end{document}